\documentstyle[12pt]{article}
\begin{document}
\begin{titlepage}

\title{ Canonical Formulation of Gravitational Teleparallelism
in 2+1 Dimensions in Schwinger's Time Gauge}
\author{A. A. Sousa and J. W. Maluf$\,^{*}$ \\ 
C. P. 04385 \\
Instituto de F\'isica, Universidade de Bras\'ilia\\
70.919-970 Bras\'ilia DF\\
Brazil\\}

\maketitle

\begin{abstract}
We consider the most general class of teleparallel gravitational
{}{}theories quadratic in the torsion tensor, in three
space-time dimensions, and carry out a detailed
investigation of its Hamiltonian formulation in Schwinger's
time gauge. This general class is given by a family
of three-parameter theories. A consistent implementation of the
Legendre transform reduces the original
theory to a one-parameter family of theories. By calculating
Poisson brackets we show explicitly
that the constraints of the theory constitute a first-class set.
Therefore the resulting theory is well defined with regard to
time evolution.  
The structure of the Hamiltonian theory rules out the existence
of the Newtonian limit.
\end{abstract}

\vfill

\noindent (*) e-mail: wadih@fis.unb.br
\end{titlepage}
\newpage

\noindent {\bf \S 1. Introduction}\par
\bigskip

Gravitational theories in three space-time dimensions have
attracted considerable attention in the last years\cite{deser}.
In particular quantum effects in this simplified geometrical
context were investigated (see, for instance,
references\cite{Achuarro,carlip}).
The hope is that lower dimensional theories would provide hints as
to the quantization of four-dimensional general relativity.
It is known that vacuum Einstein's
theory in 2+1 dimensions does not yield a suitable description
of the gravitational field\cite{giddings,deser2}. Since in 2+1
dimensions the Riemann and Ricci tensors have the same number of
components, the vanishing of Einstein's equations imply that the
full curvature tensor vanishes as well. Therefore the source-free
space-time is flat, and thus the existence of black holes is
prevented. It would be interesting to find a theoretical
formulation of 2+1 general relativity that display two important
features: the Newtonian limit and a black hole solution.

General relativity can be described in the alternative framework
of the teleparallel geometry.
A four-dimensional formulation of gravitational
teleparallel theories, quadratic in the torsion tensor, and
formulated with arbitrary parameters was proposed
by Hayashi and Shirafuji\cite{hayashi}.
This formulation was successful in that the desirable features
above were obtained for a large class of theories.
A canonical formulation of the teleparallel {\it equivalent}
of general relativity (TEGR) was developed in \cite{maluf 94}.

In this paper we consider an arbitrary teleparallel theory of
gravity in 2+1 dimensions, expressed by a three-parameter
family of theories,  and show that by means of a consistent
Legendre transform this arbitrary theory reduces to a
one-parameter family of theories. The 2+1 decomposition is
carried out in Schwinger's time gauge\cite{schwinger}. Moreover,
we calculate all relevant Poisson brackets and conclude that the
constraints of the theory are first class. As we will show, the
2+1 constraint algebra differs slightly from the previously
evaluated algebra in 3+1 dimensions\cite{maluf 94}.
The resulting
theory shares similarities with the Hamiltonian formulation
of the four-dimensional teleparallel
equivalent of general relativity\cite{maluf 94}.

The main motivation for considering the TEGR is that the energy
and momentum of the gravitational field can be interpreted as
arising from the constraint equations of the theory\cite{maluf 95}.   
All analysis carried out so far indicate that the gravitational
energy is consistently defined by means of the expression
that arises in the realm of the TEGR. The most relevant and
successful application amounts to the evaluation of the
irreducible mass of rotating black holes\cite{maluf 96}.
Recently the loss of mass by means of gravitational waves in the
context of Bondi's radiating metric has been
investigated\cite{maluf 99}.

A family of three-parameter teleparallel theories in 2+1 dimensions,
in the Lagrangian formulation,  was proposed and investigated by
Kawai\cite{kawai1}. By establishing conditions on the parameters,
black hole solutions were obtained\cite{kawai1,kawai2}. Such black
hole solutions are quite different from the Schwarzschild and Kerr
black holes in 3+1 dimensions. However, Kawai did not consider the
cosmological constant in the theory.

The arbitrary teleparallel theory we address corresponds
precisely to Kawai's formulation. We have constructed the canonical
formulation of the latter by applying Dirac's formalism for
constrained Hamiltonian systems\cite{dirac}.
Therefore we arrive at a one-parameter class of theories by
only requiring it to have a well defined Hamiltonian formulation.

The paper is divided as follows. In \S 2 we introduce the
Lagrangian formulation. The 2+1 space-time decomposition and the
canonical formalism is developed in \S 3. In this section
we provide the details of the Legendre transform. The constraint
algebra is presented in \S 4. The calculations that yield the
constraint algebra are too intricate, and therefore we have omitted
it here. In \S 5 we show that the conditions obtained on the
parameters of the theory rule out the existence of the Newtonian
limit. Finally in \S 6 we present our conclusions.\

\noindent Our notation is the following: space-time indices
$\mu, \nu, ...$ and global SO(2,1) indices $a, b, ...$ run from
0 to 2. In the 2+1 canonical decomposition Latin indices from the
middle of the alphabet indicate space indices according to
$\mu = 0, i$, $a = (0), (i)$. The flat space-time metric is
fixed by $\eta_{(0)(0)}=-1$.\
\bigskip
\bigskip

\noindent {\bf \S 2. Lagrangian formulation of teleparallelism}\par
\noindent {\bf in (2+1) dimensions}\par
\bigskip


We begin by stating the four basic postulates that the Lagrangian
density for the gravitational field in the empty space, in the
teleparallel geometry, must satisfy. It must be invariant under
(i) coordinate transformations, (ii) global Lorentz (SO(2,1))
transformations, (iii) parity transformations, and (iv) must be
quadratic in the torsion tensor.
The most general Lagrangian density is
constructed out of triads $e^a\,_\mu$, and is given
by\cite{kawai1}

$$L_{0}=e\,(c_{1}t^{abc}t_{abc}+c_{2}v^{a}v_{a}+c_{3}a_{abc}a^{abc})
\eqno(2.1)$$

where $c_{1}$, $c_{2}$ and $c_{3}$ are constants and

$$t_{abc}=\frac{1}{2}\left( T_{abc}+T_{bac}\right)
+\frac{1}{4}\left( \eta
_{ca}v_{b}+\eta _{cb}v_{a}\right) -\frac{1}{2}\eta _{ab}v_{c}{ \ \ ,}
\eqno(2.2)$$

$$v_{a}=T^{b}{}_{ba}\equiv T_a{ \ \ \ ,}
\eqno(2.3)$$

$$
a_{abc}=\frac{1}{3}\left( T_{abc}+T_{cab}+T_{bca}\right) { \ \ ,}
\eqno(2.4)$$

$$
T_{abc}=e_{b}{}^{\mu }e_{c}{}^{\nu }T_{a\mu \nu }
=e_b\,^\mu e_c\,^\nu(\partial_\mu e_{a\nu}-\partial_\nu e_{a\mu})
{ \ \ .} \eqno(2.5)$$

\noindent where $e={\tt det}(e^a\,_\mu)$.

The Lagrangian density (2.1) is constructed out of the
anti-symmetric part of the connection $\Gamma^\lambda_{\mu\nu}
=e^{a\lambda}\partial_\mu e_{a\nu}$, whose curvature tensor
vanishes identically. Such connection defines a space with
absolute parallelism, or teleparallelism.

The definitions above correspond to the irreducible components of
the torsion tensor. In order to construct the Hamiltonian
formulation we need to rewrite $L_0$ such that the torsion tensor
appears as a multiplicative quantity. It can be shown that $L_0$ can
be rewritten as

$$
L_{0}=e\,(c_{1}X^{abc}T_{abc}+c_{2}Y^{abc}T_{abc}+
c_{3}Z^{abc}T_{abc})
{ \ \ ,}\eqno(2.6)$$
where

$$
X^{abc}=\frac{1}{2}T^{abc}+\frac{1}{4}T^{bac}-
\frac{1}{4}T^{cab}+\frac{3}{8}%
\left( \eta ^{ca}v^{b}-\eta ^{ba}v^{c}\right) { \ \ \ ,}
\eqno(2.7)$$

$$
Y^{abc}=\frac{1}{2}\left( \eta ^{ab}v^{c}
-\eta ^{ac}v^{b}\right) { \ \ ,%
}\eqno(2.8)$$

$$
Z^{abc}=\frac{1}{3}\left( T^{abc}+T^{bca}+T^{cab}\right) .
\eqno(2.9)$$
The definitions above satisfy

$$
X^{abc}=-X^{acb};Y^{abc}=-Y^{acb};Z^{abc}=-Z^{acb}{ \ \ \ ,}
\eqno(2.10)$$

The quantities $X^{abc},$ $Y^{abc},$ $Z^{abc}$ have altogether the
same number of components of $T^{abc}$. It can be verified that

$$
X^{abc}+X^{bca}+X^{cab}\equiv 0{ \ \ .}
\eqno(2.11)$$

\noindent Let us define $\Sigma^{abc}$ by

$$
\Sigma^{abc}=c_{1}X^{abc}+c_{2}Y^{abc}+c_{3}Z^{abc}\;.  
\eqno(2.12)$$

\noindent In terms of $\Sigma^{abc}$, $L_0$ can be
simply written as

$$
L_{0}=e\Sigma^{abc}T_{abc}{ \ \ .}  
\eqno(2.13)$$

In order to carry out the Hamiltonian formulation we need to
write the Lagrangian density in a Palatini-type Lagrangian
density. The latter is achieved by introducing the field
variable $\Delta_{abc}=-\Delta_{acb}$,
that will be ultimately identified with the torsion tensor by means
of the field equations. By following the procedure of \cite{maluf 94}
we write the first order differential form of $L_0$ as

$$
L\left( e_{a\mu },\Delta _{abc}\right) =-e\left( c_{1}\Theta
^{abc}+c_{2}\Omega ^{abc}+c_{3}\Gamma ^{abc}\right) \left( \Delta
_{abc}-2T_{abc}\right) { \ \ ,}  
\eqno(2.14)$$
where
$\Theta ^{abc},$ $\Omega ^{abc}$ and $\Gamma ^{abc}$ are defined
in similarity to $X^{abc}$, $Y^{abc}$ and $Z^{abc}$, respectively:

$$
\Theta ^{abc}=
\frac{1}{2}\Delta ^{abc}+\frac{1}{4}\Delta ^{bac}-\frac{1}{4}%
\Delta ^{cab}+\frac{3}{8}\left( \eta ^{ca}\Delta ^{b}-\eta ^{ba}\Delta
^{c}\right) { \ \ ,}
\eqno(2.15)$$

$$
\Omega ^{abc}=
\frac{1}{2}\left( \eta ^{ab}\Delta ^{c}-\eta ^{ac}\Delta
^{b}\right) { \ \ ,}
\eqno(2.16)$$

$$
\Gamma ^{abc}=\frac{1}{3}\left( \Delta ^{abc}+\Delta ^{bca}+\Delta
^{cab}\right) { \ \ .}
\eqno(2.17)$$
The three quantities above are anti-symmetric in the last two indices.
The quantity $\Delta^b$ is defined by $\Delta^b=\Delta^a\,_a\,^b$.

The field equations are most easily obtained by making use of the
following three identities:

$$
X^{abc}\Delta _{abc}=\Theta ^{abc}T_{abc}{ \ \ ,}
\eqno(2.18)$$

$$
Y^{abc}\Delta _{abc}=\Omega ^{abc}T_{abc}{ \ \ ,}
\eqno(2.19)$$

$$
Z^{abc}\Delta _{abc}=\Gamma ^{abc}T_{abc}{ \ \ .}
\eqno(2.20)$$

By taking into account the identities above the Lagrangian density
in first order formalism is identically rewritten as

$$
L =-ec_{1}\Theta ^{abc}\Delta _{abc}+2ec_{1}X^{abc}\Delta
_{abc}-ec_{2}\Omega ^{abc}\Delta _{abc}+2ec_{2}Y^{abc}\Delta _{abc}$$

$$-ec_{3}\Gamma ^{abc}\Delta _{abc}+2ec_{3}Z^{abc}\Delta _{abc}
\;.\eqno(2.21)$$

The variation of $L$ with respect to $\Delta_{def}$ is given by

$$
\frac{\delta L}{\delta \Delta _{def}} =-ec_{1}\frac{\delta \left( \Theta
^{abc}\Delta _{abc}\right) }{\delta \Delta _{def}}+2ec_{1}X^{abc}\frac{%
\delta \left( \Delta _{abc}\right) }{\delta \Delta _{def}}$$

$$
-ec_{2}\frac{\delta \left( \Omega ^{abc}\Delta _{abc}\right) }{\delta
\Delta _{def}}+2ec_{2}Y^{abc}\frac{\delta \left( \Delta _{abc}\right) }{%
\delta \Delta _{def}}$$

$$-ec_{3}\frac{\delta \left( \Gamma ^{abc}\Delta _{abc}\right) }{\delta
\Delta _{def}}+
2ec_{3}Z^{abc}\frac{\delta \left( \Delta _{abc}\right) }{%
\delta \Delta _{def}},\eqno(2.22)$$

$$
\frac{\delta L}{\delta \Delta _{def}} =-2ec_{1}\Theta
^{def}+2ec_{1}X^{def}-2ec_{2}\Omega ^{def}$$

$$+2ec_{2}Y^{def}-2ec_{3}\Gamma ^{def}+2ec_{3}Z^{def},
\eqno(2.23)$$
where we have used

$$
\frac{\delta \left( \Theta ^{abc}\Delta _{abc}\right) }
{\delta \Delta _{def}}
=2\Theta ^{abc}\frac{\delta \Delta _{abc}}
{\delta \Delta _{def}}\;,\eqno(2.24)$$

$$\frac{\delta \left(
\Omega ^{abc}\Delta _{abc}\right) }{\delta \Delta _{def}}
=2\Omega ^{abc}\frac{\delta \Delta _{abc}}{\delta \Delta _{def}}\;,
\eqno(2.25)$$

$$\frac{\delta \left(
\Gamma ^{abc}\Delta _{abc}\right) }{\delta \Delta _{def}}
=2\Gamma ^{abc}\frac{\delta \Delta _{abc}}{\delta \Delta _{def}}\;,
\eqno(2.26)$$
since the quantities between parentheses in the left hand side of
the expressions above are quadratic in $\Delta _{abc}$.
This result can be verified by explicit calculations.

By considering the action integral

$$
I=\int L{ }d^{3}x,
\eqno(2.27)$$
and imposing the variation

$$
\frac{\delta I}{\delta \Delta _{def}}=
\int \frac{\delta L}{\delta \Delta
_{def}}{ }d^{3}x=0, 
\eqno(2.28)$$
we arrive at

$$
c_{1}\left( X^{def}-\Theta ^{def}\right) +c_{2}\left( Y^{def}-\Omega
^{def}\right) +c_{3}\left( Z^{def}-\Gamma ^{def}\right) =0. 
\eqno(2.29)$$

\noindent By taking into account equations $(2.7)\sim(2.9)$ and
$(2.15)\sim(2.17)$, the only solution to equation (2.29),
for arbitrary values of $c_{i}$, is given by

$$
\Delta _{abc}=T_{abc}= e_{b}{}^{\mu }e_c\,^{\nu }T_{a\mu \nu }  \;,
\eqno(2.30)$$
that implies

$$
X^{abc}=\Theta ^{abc}\;,
\eqno(2.31)$$

$$
Y^{abc}=\Omega ^{abc}\;,
\eqno(2.32)$$

$$
Z^{abc}=\Gamma ^{abc}\;.
\eqno(2.33)$$
Note that the equation (2.29) represents nine equations
for nine unknown quantities $\Delta _{abc}$.\par
\bigskip
\bigskip

\noindent {\bf \S 3. The Hamiltonian formulation}\par
\bigskip


In this section it will be necessary to make a change of notation.
The three-dimensional space-time triads of the last section will
be denoted here as $^{3}e^{a}{}_{\mu }$, whereas diads restricted
to the two-dimensional spacelike surface will be represented simply
by $e^{a}{}_{\mu }$. This distinction is mandatory in the 2+1
decomposition of the triads. The latter is similar to the 3+1
decomposition of tetrad fields.

The 2+1 decomposition of triads is given by

\[
^{3}e^{a}{}_{k}=e^{a}{}_{k}
\]

\[
^{3}e^{ai}=e^{ai}+\frac{N^{i}}{N}\eta ^{a}
\]

\[
e^{ai}=\overline{g}^{ik{ }}e^{a}{}_{k}\hspace{0in}\hspace{0in}\hspace{%
0in}{ \hspace{0pt} \ \ \ \ \ }\eta ^{a}=-N^{{ \ \ }3}e^{a0}
\]

\[
^{3}e^{a}{}_{0}=N^{i}e^{a}{}_{i}+N\eta ^{a}
\]

$$
\eta ^{a}e_{a}{}_{k}=0{ \ \ \ \ \ \ \ \ \ }\eta _{a}\eta ^{a}=-1{
\ \ \ \ \ \ \ \ \ }^{3}e=Ne 
\eqno(3.1)$$
where $N$ and $N^i$ are the lapse and shift functions, respectively,
and

\[
^{3}e=\det \left( ^{3}e^{a}{}_{\mu }\right)
\]

\[
g_{ij}=e^{a}{}_{i}e_{aj}
\]

$$
\overline{g}^{ij}g_{jk}=\delta _{k}^{i}\;.
\eqno(3.2)$$
Therefore,

$$
\left( \overline{g}^{ik}\right) \sim \left( g_{ij}\right) ^{-1}\;.
\eqno(3.3)$$
It follows that

\[
e^{bk}e_{bj}=\delta _{j}^{k},
\]

$$
e^{a}{}_{i}e^{bi}=\eta^{ab}+\eta ^{a}\eta ^{b}.
\eqno(3.4)$$
The components $e^{ai}$ and $e^{a}{}_{k}\hspace{0in}$ are now
restricted to the two-dimensional spacelike surface.

The Hamiltonian formulation is achieved by rewriting
the Lagrangian density (2.14) in the form
$L=p\dot q-H$.
For this purpose we define, 
in analogy with (2.12), the
field quantity $\Lambda^{abc}$ according to 

$$
\Lambda ^{abc}=c_{1}\Theta ^{abc}+
c_{2}\Omega ^{abc}+c_{3}\Gamma ^{abc}{
\ \ ,}  
\eqno(3.5)$$

By means of (3.5) we define $P^{ai}$, the momentum canonically
conjugated to $e_{ai}$. It is given by

$$
P^{ai}=4\,^3e\Lambda ^{a0i}=4
ee_{b}{}^{i}\eta _{c}\Lambda ^{abc}{ \ \ \ .}
\eqno(3.6)$$

In terms of $P^{ai}$ we write the Lagrangian density as

$$
L =P^{ai}\stackrel{.}{e}_{ai}+^{3}e_{a0}\partial _{i}P^{ai}$$

$$+2Ne\Lambda ^{aij}T_{aij}+N^{k}P^{ai}T_{aik}$$

$$-Ne\Lambda ^{abc}\Delta _{abc}
-\partial _{i}\left[\, ^3e_{a0} P^{ai}\right] { \ \ .}
\eqno(3.7)$$
We note that there is no time derivative of $^3e_{a0}$. Thus the
momentum $P^{a0}$ canonically conjugated to $e_{a0}$ is taken to
vanish from the outset.

At this point we adopt Schwinger's time gauge \cite{schwinger},

$$
\eta ^{a}=\delta _{(0)}^{a}{ \ \ \ \ \ \ , \ \ \ \ \ \ }\eta
_{a}=-\delta _{a}^{(0)}{ \ \ .} 
\eqno(3.8)$$

\noindent Conditions above imply

$$^3e_{(k)}\,^0= e^{(0)}\,_i=0\;.$$

\noindent The time gauge is taken to hold {\it before} varying
the action, since the fixing of this gauge is not a consequence
of any local symmetry of the theory.  The imposition of
time gauge actually corresponds to a reduction in the
configuration space of the theory. Whereas the fixation of this
gauge in the Lagrangian formulation of a teleparallel theory
involves some intrincacies, in the Hamiltonian formulation it
amounts to a straightforward procedure.

By making use of the Lagrangian field equations we identify

$$
\Delta _{aij }=T_{aij } \eqno(3.9)$$
in $L$, since these equations do not involve time derivatives.

Following \cite{maluf 94}, we wish to establish a 2+1 decomposition
for $\Lambda^{abc}$ that distinguishes the components of the latter
that are projected (restricted) to the two-dimensional spacelike
surface from those that define the canonical momenta. Assuming
Schwinger's time gauge we write

$$
\Lambda ^{abc}=\frac{1}{4e}\left( \eta ^{b}e^{c}{ }_{i}P^{ai}-\eta
^{c}e^{b}{ }_{i}P^{ai}\right) +e^{b}{ }_{i}e^{c}{ }%
_{j}\Lambda ^{aij}{ \ \ ,} \eqno(3.10)$$
where

$$
\Lambda ^{aij}=e^{a}{ }_{k}\Lambda ^{kij}{ \hspace{0in} .}
\eqno(3.11)$$

\noindent In the expression above $\Lambda^{kij}$ is a tensor on
the two-dimensional spacelike surface.

The Legendre transform would be straightforward if $\Lambda^{aij}$
would depend only on $e_{(k)j}$ and its spatial derivatives. However
in general
this is not the case. The main issue is that after the Legendre
transform has been performed, the Hamiltonian density cannot depend
on the ``velocities" $\Delta _{a0j}=T_{a0j}$, which
contain terms of the type $\stackrel{.}{e}_{ai}$.
The quantities $\Lambda^{aij}$ in (3.10) in general contain terms
like $\Delta_{a0j}$. 
Such terms cannot be present in the final form
of the Hamiltonian density obtained via
$H=P^{ai}(x)\dot e_{ai}(x) -L$. This goal
will be achieved by posing restrictions on the constants $c_i$,
as we will see, and by invoking the time gauge condition.

Since the momenta is defined by (3.6), and since $P^{ai}$
is an {\it irreducible} component of $\Lambda^{abc}$ as
given by (3.10), we expect the contribution of (3.11)
to the Lagrangian density not to yield
velocities terms of the type $\Delta_{a0j}$.
Thus $\Lambda^{aij}$ in (3.10) must not
depend on the momenta, and therefore it cannot lead to
the emergence of velocities in the final form of $L$.

As we mentioned earlier,
the time gauge condition reduces the configuration space of the
theory from the SO(2,1) (in $L$) to the SO(2) group (in $H$). As
a consequence of $\dot e_{(0)i}=0$ the teleparallel geometry is
restricted to the two-dimensional  spacelike surface.\\

\bigskip

We will now express the several components of $L$ in (3.7) by means
of the 2+1 decomposition of the triads and of $\Lambda^{abc}$.
Considering definitions (3.5) and (3.6) we can obtain by explicit
calculations the expression of $P^{(0)k}$. It is given by

$$
P^{\left( 0\right) k}=-2eT^{\left( 0\right) }{}_{\left( 0\right)
}{}^{k}\left( \frac{3}{4}c_{1}+c_{2}\right)
+eT^{k}(\frac{3}{2}c_{1}-2c_{2})%
{ \ \ .}  \label{7}
\eqno(3.12)$$
where $T^k=T^{(i)}\,_{(i)}\,^k$.

Considering first $^3e_{a0}\partial_i P^{ai}$ we find

$$
^3e_{a0}\partial _{i}P^{ai}=
N^{k}e_{\left( j\right) k}\partial _{i}P^{\left(
j\right) i}-N\partial _{k}P^{\left( 0\right) k}{ \ \ .}
\eqno(3.13)$$

As to the surfaced term $-\partial_i(\,^3e_{a0}P^{ai})$ we have

$$
-\partial _{i}\left[ P^{ai}\left( N^{k}e_{ak}+N\eta _{a}\right) \right]
=-\partial _{i}\left[ N_{k}P^{ki}\right] +\partial _{i}\left[ NP^{\left(
0\right) i}\right] { \ \ .}
\eqno(3.14)$$

Let us consider the term  $-Ne\Lambda ^{abc}\Delta _{abc}$.
By means of (3.1), (3.8) and (3.10) this term can be rewritten
after a long calculation as

$$
-N\,e\,\Lambda^{abc}\Delta_{abc}  =
+N\left( \frac{c_{1}}{4}-\frac{c_{3}}{3}\right) ^{-1}\times $$

$$\times\biggl\{\frac{1}{16e}\left( P^{ij}P_{ji}-
P^{\left( 0\right) l}P^{\left(
0\right) }{ }_{l}\right)$$

$$+\frac{1}{2}\left( e_{\left( m\right) i}P^{\left( m\right) }{ }%
_{j}{}\Lambda ^{\left( 0\right) ij}\right)$$

$$-ee^{\left( m\right) }{ }_{i}e_{\left( n\right) }{ }^{k}\Lambda
^{\left( n\right) ij}\Lambda _{\left( m\right) kj}$$

$$+\left( \frac{3c_{1}}{8c_{2}}-\frac{1}{2}\right)%
\{\frac{1}{16e}\left[ P^{2}-P^{\left( 0\right) l}P^{\left(0\right)
}{ }_{l}\right]$$

$$+\frac{1}{2}P^{\left( 0\right) }{ }_{k}e_{\left( m\right) j}\Lambda
^{\left( m\right) jk}$$

$$-ee_{\left( m\right) i}\Lambda ^{\left( m\right) ik}e_{\left(
n\right) j}\Lambda ^{\left( n\right) j}{ }_{k}\}$$

$$+\left( \frac{c_{1}}{4}+\frac{2c_{3}}{3}\right) \{\frac{1}{4}\Delta
_{ij\left( 0\right) }P^{ji}-e\Delta _{i\left( 0\right) j}\Lambda ^{\left(
0\right) ij}$$

$$+e\Delta _{ikj}\Lambda ^{kij}+\Delta _{\left( 0\right) \left( 0\right)
i}P^{\left( 0\right) i}
+\Delta _{\left( 0\right) ij}P^{ij}\}\biggr\}
\eqno(3.15)$$

\noindent Space indices are raised and lowered by means of
$e_{(i)j}$ and $e^{(k)l}$. The quantity $P$ is defined by
$P=P^{(i)j}e_{(i)j}$.
We observe that $\Lambda^{(0)ij}$ contains time derivatives
$\dot e_{(i)j}$. However, we note that the third term
on the right hand side of the
expression above can be rewritten as

$$
{1\over 2}(e_{(m)i}P^{(m)}\,_j\Lambda^{(0)ij})=
{1\over 2}(P_{\lbrack ij \rbrack}\Lambda^{(0)ij})\;,
\eqno(3.16)$$
where $\lbrack .. \rbrack$ denotes anti-symmetrization. It is
known that in tetrad type theories of gravity the anti-symmetric
part of the canonical momenta vanishes weakly. Let us obtain here
the full expression of $P_{\lbrack ij \rbrack}$. Making use of
(3.5) and  (3.6) we find

$$
P_{\left[ ij\right] }+
eT_{\left( 0\right) ij}\left( c_{1}-\frac{4}{3}%
c_{3}\right)
+eT_{\left[ i\left| \left( 0\right) \right| j\right] }\left(
c_{1}+\frac{8}{3}c_{3}\right) =0{ \ \ .}
\eqno(3.17)$$
Note that in the time gauge the term $T_{\left( 0\right) ij}$
vanishes.

We consider finally the term $2Ne\Lambda ^{aij}T_{aij}$. In the
2+1 decomposition it reads

$$
2Ne\Lambda ^{aij}T_{aij}=
2Ne(\Lambda ^{\left( 0\right) ij}T_{\left(
0\right) ij}+
\Lambda ^{\left( k\right) ij}T_{\left( k\right) ij}).  \nonumber
\eqno(3.18)$$

\noindent The time gauge simplifies the expression above in two
aspects. Because of it, the first term on the right hand side
vanishes. Moreover, it can be verified by explicit calculations
that the second term does not contain velocity terms
$\Delta _{a0j}=T_{a0j}$.

By carefully inspecting expressions (3.12) and (3.15) we arrive
at the conditions that allow a well defined Legendre transform.
From (3.12) we observe that we must demand

$$
c_{2}+\frac{3c_{1}}{4}=0\,.\eqno(3.19)$$
Furthermore by requiring

$$
c_{1}+\frac{8c_{3}}{3}=0\,,\eqno(3.20)$$

\noindent the last five terms of expression (3.15) drop out.
Four of these terms are velocity dependent.
In view of the argument presented above, according to which
the momentum $P^{ai}$ is an irreducible component of
$\Lambda^{abc}$, these terms must be eliminated
in the Legendre transform. As a consequence, expression
(3.17) also becomes exempt of velocity terms. In the time
gauge equation (3.17) reads

$$
P_{\left[ ij\right] }=0\,,\eqno(3.21)$$
and therefore it must appear in the Hamiltonian formulation as
constraint equations.

We remark that the imposition of (3.19) implies the elimination of
the ``velocity" $T_{(0)0k}$ from the expression of $P^{(0)i}$.
However, in view of the time gauge condition such term does not
contain any time derivative. It contains only a spatial
derivative of the lapse function. We know that the variation
of the Hamiltonian with respect to the lapse function yields
the Hamiltonian constraint. It turns out that  we still
have to require (3.19), otherwise the presence of such derivative
of the lapse function would render a very complicated expression
for the Hamiltonian constraint, which does not satisfy the
constraint algebra to be presented in the next section. Moreover
the lapse function would no longer play the role of a genuine
Lagrange multiplier.

The actual necessity of demanding condition (3.19) will
be demonstrated in \S 4. It will be shown that without
this condition the Hamiltonian formulation cannot be
consistently established.

We can write the final form of $L$ by collecting the remaining
terms. We choose to write the Lagrangian density in terms of
$c_1$ only. By factorizing the lapse and shift functions we obtain

$$
L =P^{\left( j\right) i}\stackrel{.}{e}_{\left( j\right)
i}+N^{k}C_{k}+NC-\partial _{i}\left[ N_{k}P^{ki}+N(3c_{1}eT^{i})\right]
+\lambda ^{ij}P_{\left[ ij\right] }\eqno(3.22)$$
where $\lambda^{ij}$ are Lagrange multipliers.
The Hamiltonian and vector constraints are given respectively by

$$
C=\frac{1}{6ec_{1}}\left( P^{ij}P_{ji}-P^{2}\right) +eT^{ikj}\Sigma
_{ikj}-\partial _{k}\left[ 3c_{1}eT^{k}\right] { \
\ ,}\eqno(3.23)$$

$$
C_{k}=e_{\left( j\right) k}\partial _{i}P^{\left( j\right) i}+P^{\left(
j\right) i}T_{\left( j\right) ik}.\eqno(3.24)$$
We remark that $\Sigma_{kij}$ that appears in $C$ is a function of
$e_{(i)j}$ and its spatial derivatives only. The Lagrange multipliers
$\lambda^{ij}$ are ultimately determined by evaluating the field
equation $\dot e_{(i)j}(x)=\lbrace e_{(i)j}(x), {\cal H}\rbrace$,
where ${\cal H}$ is the total Hamiltonian, and by imposing
$P_{\lbrack ij \rbrack}=0$.

Therefore we conclude this section by observing that the
imposition of a well defined Legendre transform has reduced
the three-parameter to a one-parameter family of theories.\par
\bigskip
\bigskip

\noindent {\bf \S 4. The constraint algebra}\par

\bigskip

A consistent implementation of the constraint algebra is a necessary
condition for the Hamiltonian formulation. However, it is not
a sufficient condition. It remains to be verified whether the
constraint structure of the theory is consistently implemented
in the sense of Dirac's formulation of constrained Hamiltonian
systems. We recall that 
the Hamiltonian formulation of the TEGR is determined
by a set of first class constraints\cite{maluf 94}.

The Hamiltonian formulation determined by $(3.21)\sim(3.24)$
is very much
similar to the 3+1 Hamiltonian formulation of the TEGR, the
difference residing in the presence of the constant $c_1$ in
$(3.22)\sim(3.24)$ and in the numerical coefficient of the
$P^2$ term in the Hamiltonian constraint $C$.

The calculations that lead to the constraint algebra between
(3.21), (3.23) and (3.24) are extremely long and intricate. Here we
just provide the final expressions. Regardless of the value of
$c_1$ the constraint algebra ``closes", and therefore (3.21), (3.23)
and (3.24) constitute a first class set. Except for the numerical
value of the contraction of the metric tensor with itself, the
calculations in 2+1 are almost identical to the calculations in
the 3+1 formulation.

The constraint algebra is given by

$$
\left\{ C(x),C(y)\right\}
=\biggl[ -g^{ik}(x)C_{i}(x)-2\partial_{j}P^{\left[ jk\right] }(x)$$

$$+P_{\left[ mn\right] }\left( -\frac{1}{2}%
T^{kmn}+T^{mnk}\right) \biggr] 
\frac{\partial }{\partial x^{k}}\delta \left( x-y\right) -\left(
x\leftrightarrow y\right) ,
\eqno(4.1)$$

$$
\left\{ C(x),C_{k}(y)\right\}
=C(y)\frac{\partial }{\partial y^{k}}\delta
\left( x-y\right) ,
\eqno(4.2)$$

$$
\left\{ C_{j}(x),C_{k}(y)\right\} =-
C_{k}(x)\frac{\partial }{\partial x^{j}}%
\delta \left( x-y\right) +
C_{j}(y)\frac{\partial }{\partial y^{k}}\delta
\left( x-y\right) .
\eqno(4.3)$$
Moreover we have

$$
\left\{C(x),P^{\left[\left( m\right) \left( n\right)\right]}(y)\right\}
=\left\{ C_{k}(x),P^{\left[ \left( m\right) \left( n\right) \right]
}(y)\right\} =0
\eqno(4.4)$$

$$
\left\{ P^{\left[ \left( m\right) \left( n\right) \right] }(x),P^{\left[
\left( i\right) \left( j\right) \right] }(y)\right\}
=\biggl(\eta ^{\left(
n\right) \left( j\right) }P^{\left[ \left( m\right) \left( i\right) \right]
}(x)-\eta ^{\left( n\right) \left( i\right) }
P^{\left[ \left( m\right) \left(
j\right) \right] }(x)$$

$$
+\eta ^{\left( m\right) \left( i\right) }P^{\left[ \left( n\right) \left(
j\right) \right] }(x)-
\eta ^{\left( m\right) \left( j\right) }P^{\left[ \left( n\right) \left(
i\right) \right]}(x)\biggr) \delta \left( x-y\right)
\eqno(4.5)$$
Note that by making
$P^{\left[ \left( m\right) \left( n\right) \right] }=0$,
we obtain the constraint algebra of the 2+1 ADM formulation.
The value of $c_1$ remains arbitrary.

We are now in a position to discuss the necessity of (3.19).
By not requiring (3.19) the expression
of $P^{(0)k}$, given by (3.12), acquires an extra term given by

$$-2eT^{(0)}\,_{(0)}\,^k\biggl( {3\over 4}c_1+c_2\biggr)=
2e\biggl({3\over 4}c_1+c_2\biggr)g^{kl}{1\over N}\partial_l N\;.$$

\noindent As a consequence the Hamiltonian density acquires an
extra term as well. Considering equation (3.13) this extra term 
reads

$$W=-2eN\biggl({3\over 4}c_1+c_2\biggr)\partial_k\biggl(
eg^{kl}{1\over N}\partial_l N\biggr)\;.$$

\noindent It is easy to see that this extra term spoils the
closure of the constraint algebra. The simplest way of observing
the emergence of troublesome terms in the constraint algebra of
the theory is by verifying the consistency of the vector constraint
$C_k$. We remark that the expression of $C_k$ does not depend on
the imposition of condition (3.19). Therefore this condition is
not {\it a priori} assumed. 
By calculating the time evolution $\dot C_k(x)=
\lbrace C_k(x), {\cal H}\rbrace$, where ${\cal H}$ is the total
Hamiltonian, we must evaluate

$$\int d^3y\lbrace C_k(x), W(y)\rbrace=$$

$$-2e\biggl(c_2+{3\over 4}c_1\biggr){1\over N}
(\partial_iN)(\partial_j N)(e^{al}g^{ij}-2g^{il}e^{aj})
(\partial_l e_{ak}+\partial_k e_{al})\;.$$

\noindent We cannot take the right hand side of the expression
above as a constraint (either first class or second class)
in conjunction with the additional terms in the expression
of $\dot C_k(x)$, 
otherwise several other constraints would
emerge by means of consistency conditions, and eventually all
degrees of freedom would be exhausted. Therefore condition
(3.19) is mandatory.\par
\bigskip
\bigskip

\noindent {\bf \S 5. The absence of a Newtonian limit}\par
\bigskip


The existence of the Newtonian limit was investigated in ref.
\cite{kawai1} by considering static fields with circular symmetry,
in the absence of spinorial particles and without cosmological
constant. It was found a relation that leads to the Newtonian
limit and that in our notation reads

$$
3c_{1}+4c_{2} =-6\,c_{1}c_{2}\;,$$

$$c_{1}c_{2} \neq 0.\eqno(5.1)$$

Conditions (3.19) and (3.20) violate conditions
above for the existence of a Newtonian limit. Therefore a well
defined theory from the point of view of the initial value
problem cannot display such limit.

The field equations for the Lagrangian density (2.1) were obtained in
ref. \cite{kawai1}. It was noticed that the field equations are
equivalent to Einstein's three-dimensional field equations if the
parameters satisfy

$$
c_{1}+\frac{2}{3} =0\;,$$

$$
c_{2}-\frac{1}{2} =0\;,$$

$$
c_{3}-\frac{1}{4} =0\;.\eqno(5.2)$$

By comparing the equations above with (3.19) and (3.20) we conclude
that the theory defined by $(3.22)\sim(3.24)$ is equivalent to the
source free
Einstein's general relativity in 2+1 dimensions provided we fix
$c_1=-{2\over 3}$. However, it must be noted that the interaction
type between matter (spin ${1\over 2}$) fields and the gravitational
field in teleparallel theories is different from that in
Einstein's theory (see equation (3.20') of \cite{kawai1}).\par
\bigskip
\bigskip

\noindent {\bf \S 6. Conclusions}\par
\bigskip


In this paper we have investigated the existence of a viable
theory of 2+1 dimensional gravity by only requiring it to
have a well defined Hamiltonian formulation. A consistent
implementation of the Legendre transform reduced the original
three-parameter to a one-parameter theory. The resulting
theory corresponds to a constrained Hamiltonian system with
first class constraints, with total Hamiltonian given by

$${\cal H}\;=\;
-\int d^3x\biggl(
NC+N^iC_i+\lambda^{ij}P_{ij}
-\partial_i\lbrack N_kP^{ki}+N(3c_1eT^i)\rbrack \biggr)\;.$$

The final form of the theory shares similarities with
the 3+1 canonical formulation of the teleparallel equivalent of
general relativity. The successful applicability of the Hamiltonian
formalism to lower dimensional formalisms is a positive feature of
Dirac's formulation of Hamiltonian constrained systems. Furthermore
it supports the conjecture that teleparallel theories may acquire
a prominent status in the investigation of gravity theories.

Finally we mention that a theory obtainable from the Lagrangian
density (3.22) by adding a negative cosmological constant has
the well known BTZ black hole solution\cite{btz}.
The BTZ black hole solution is
found by ascribing the free parameter $c_1$ the value
$c_1=-{2\over 3}$. This investigation will
be presented elsewhere.\par
\bigskip
\noindent {\it Acknowledgements}\par
\noindent A. A. Sousa is grateful to the Brazilian agency
CNPQ for financial support.\par

\end{document}